# Correlation between Compensation Temperatures of Magnetization and Angular Momentum in GdFeCo Ferrimagnets

Yuushou Hirata[1], Duck-Ho Kim[1†], Takaya Okuno[1], Tomoe Nishimura[1], Dae-Yun Kim[2], Yasuhiro Futakawa[3], Hiroki Yoshikawa[3], Arata Tsukamoto[3], Kab-Jin Kim[4], Sug-Bong Choe[2], and Teruo Ono[1,5†]

[1]Institute for Chemical Research, Kyoto University, Uji, Kyoto 611-0011, Japan.

[2]Department of Physics and Institute of Applied Physics, Seoul National University, Seoul 08826, Republic of Korea.

[3]College of Science and Technology, Nihon University, Funabashi, Chiba 274-8501, Japan.

[4]Department of Physics, Korea Advanced Institute of Science and Technology, Daejeon 34141, Republic of Korea.

[5]Center for Spintronics Research Network, Graduate School of Engineering Science, Osaka University, Machikaneyama 1-3, Toyonaka, Osaka 560-8531, Japan.

[†]Correspondence to: kim.duckho.23z@st.kyoto-u.ac.jp, ono@scl.kyoto-u.ac.jp

**Determining the angular momentum compensation temperature of ferrimagnets is an important step towards ferrimagnetic spintronics, but is not generally easy to achieve it experimentally. We propose a way to estimate the angular momentum compensation temperature of ferrimagnets. We find a linear relation between the compensation temperatures of the magnetization and angular momentum in GdFeCo ferrimagnetic materials, which is proved by theoretically as well as experimentally. The linearity comes from the power-law criticality and is governed by the Curie temperature and the Landé g factors of the elements composing the ferrimagnets.**

**Therefore, measuring the magnetization compensation temperature and the Curie temperature, which are easily assessable experimentally, enables to estimate the angular momentum compensation temperature of ferrimagnets. Our study provides efficient avenues into an exciting world of ferrimagnetic spintronics.**

Antiferromagnets came into the spotlight in the last decade [1–6] as a promising material for spintronics devices because they exhibit fast magnetic dynamics and low susceptibility to magnetic fields. These advantages originate from the antiferromagnetic ordering in which the magnetic moments are compensated on an atomic scale. This also implies that it is difficult to efficiently manipulate antiferromagnets using external magnetic fields, hindering the study of antiferromagnetic spin dynamics. However, recently, magnetic field-controlled antiferromagnetic spin dynamics has been achieved using ferrimagnets [7]. Hence, ferrimagnets have become a promising material in the emerging field of antiferromagnetic spintronics.

Ferrimagnetic materials comprise rare earth (RE) and transition metal (TM) compounds, wherein the spins of two inequivalent sublattices are coupled antiferromagnetically [8-10]. Because of the different Landé g factors of RE and TM elements, ferrimagnets exhibit compensation temperatures of magnetization and angular momentum [11], at which the magnetizations (angular momenta) of the RE and TM sublattices have the same magnitude but opposite directions. Consequently, the net magnetization (angular momentum) is compensated. The compensation temperatures have been studied experimentally and theoretically [7, 12–23]. In particular, Kim *et al*. recently observed fast field-driven domain wall (DW) motion in the vicinity of the compensation temperature of the angular momentum [7]. This observation reveals that ferrimagnets exhibit the antiferromagnetic dynamics because of the zero net angular momentum a t the compensation temperature of the angular momentum, even though they have magnetic moments. Although remarkable efforts have been made theoretically and experimentally [7, 12–23] in understanding the role of angular momentum compensation in DW dynamics, it is difficult to determine the angular momentum compensation temperature because of the methodological complexities, impeding the rapid development of this exciting research field as well as the fundamental understanding of the compensation temperatures.

In this letter, we report the correlation between the angular momentum and magnetization compensation temperatures in ferrimagnetic GdFeCo alloys. It is experimentally demonstrated that the angular momentum compensation temperature is directly related to the magnetization compensation temperature, regardless of the sample structures. The results show that there exists a strong correlation between the two types of compensation temperatures. We theoretically verified the correlation on the basis of a simple modeling technique. Moreover, the proposed approach is a novel method of determining the angular momentum compensation temperature.

For this study, we prepared six types of amorphous ferrimagnet GdFeCo films. Table I lists the detailed sample structures. The films were grown by co-sputtering, and the compositions were estimated from the relative deposition rates of Gd and FeCo. The samples exhibit perpendicular magnetic anisotropy (PMA) with circular domain expansion. As shown in Fig. 1(**a**), an e-beam lithography technique was applied to the structural devices with a Hall bar geometry in order to detect the anomalous Hall resistance.

First, we characterized the magnetic properties of the GdFeCo samples. Figure 1(**b**) shows the hysteresis loop of the GdFeCo microstrip. The anomalous Hall resistance $R_H$ is measured as a function of the perpendicular magnetic field $H_z$ at room temperature. The clear, square, hysteresis loop shows that the GdFeCo samples have strong PMA. The orange-colored arrow represents the magnetization switching field, which is referred to as the coercive field $H_c$.

To determine the magnetization compensation temperature $T_M$, we measured $R_H$ by sweeping $H_z$ at each temperature [7, 18, 20, 23]. The magneto-transport properties are dominated by the FeCo moment because the energy of the 4f shell of Gd is located far below the Fermi energy level [23]. A sign change in $R_H$ indicates a change in the direction of the FeCo moment. Using $R_H$ as a function of $H_z$, we define the Hall resistance difference as $\Delta R_H \equiv R_H(+H_{z,sat}) - R_H(-H_{z,sat})$ (see Fig. 1(**b**)). Figure 2(**a**) shows the Hall resistance difference $\Delta R_H$ as a function of the temperature $T$ for Sample II, where $+H_{z,sat}$ and $-H_{z,sat}$ are the saturation fields with the condition $H_c < H_{z,sat}$ (see inset of Fig. 1(**b**)). $\Delta R_H$ to zero determines $T_M = 160$ K, indicated by a blue dot.

To determine $T_A$, we measured the field-driven DW speed as a function of the temperature, as proposed elsewhere [7]. We first applied a sufficiently strong magnetic field with a magnitude of $-200$ mT ($-H_{z,sat}$) to saturate the magnetization along the $-z$ direction, and subsequently, a constant $H_z$ for driving the DW. $H_z$ is selected lower than $H_c$, to eliminate the nucleation of the domain. Next, we applied a DC current $I_x$ along the wire to detect the Hall signal (see red arrow in Fig. 1(**a**)), where $I_x$ is sufficiently small to prevent spin torques and Joule heating effect [24–26]. We then injected a current pulse $I_y$ (12 V, 100 ns) through the writing line (see blue arrow in Fig. 1(**a**)) to nucleate the reversed domain, thereby creating two DWs in the wire. The created DW moves along the wire because of the presence of $H_z$, and then, passes through the Hall bar; the DW arrival time can be detected by monitoring the change in the Hall voltage using an oscilloscope. The DW speed can be calculated from the arrival time and the distance traveled between the writing line and the Hall bar (400 μm).

Figure 2(**b**) shows the DW speed $v$ as a function of the temperature $T$ at $\mu_0 H_z = 80$ mT for Sample II. This figure clearly shows that $v$ exhibits a peak at a certain temperature (indicated by purple arrow). This tendency of $v$ with respect to $T$ is consistent with the results given elsewhere [7]. Accordingly, $T_A$ can be determined, as shown in Fig. 2(**b**). Here, the difference between $T_A$ and $T_M$ is defined as $\Delta T$ ($\equiv T_A - T_M$), indicated by black double arrows.

For a quantitative comparison, the values of $T_A$ are directly plotted with respect to $T_M$ for all the samples, as shown in Fig. 3. It is interesting to note that all the values $(T_M, T_A)$ lie on a single curve with linearity. The red line represents the best linear fit with a slope of 0.87 and a y-axis intercept of 101.1 K. From this result, we experimentally found that there exists a strong correlation between $T_M$ and $T_A$ for all the GdFeCo films.

To understand the correlation of $(T_M, T_A)$, we employ a theory based on a power-law criticality, given that the variation in the magnetization as a function of the temperature can be reasonably well approximated [27, 28, 29]. This function describes the temperature dependence of the magnetization, which can be expressed as $M(T) \sim (T_C - T)^\beta$, where $M$ is the saturation magnetization, $T_C$ is the Curie temperature, and $\beta$ is the critical exponent. Accordingly, the temperature dependencies of the magnetization for Gd and FeCo can be written as $M_{Gd}(T) = M_{Gd}(0)(1 - T/T_C)^{\beta_{Gd}}$ and $M_{FeCo}(T) = M_{FeCo}(0)(1 - T/T_C)^{\beta_{FeCo}}$, respectively, where $\beta_{Gd}$ (or $\beta_{FeCo}$) is the critical exponents of Gd (or FeCo) and $M_{Gd}(0)$ (or $M_{FeCo}(0)$) is the saturation magnetization of Gd (or FeCo) at zero temperature, where $\beta_{Gd} > \beta_{FeCo}$ and $M_{Gd}(0) > M_{FeCo}(0)$ [27, 28]. The total saturation magnetization $M_{total}$ can be determined using the relation, $M_{total}(T) = M_{FeCo}(0)(1 - T/T_C)^{\beta_{FeCo}} - M_{Gd}(0)(1 - T/T_C)^{\beta_{Gd}}$. As $M_{total} = 0$ at $T = T_M$, the following equation can be written.

$$T_C - T_M = T_C[M_{Gd}(0)/M_{FeCo}(0)]^{1/(\beta_{FeCo} - \beta_{Gd})}. \tag{1}$$

**5**

Similarly, the total angular momentum $A_{\text{total}}(T)$ can be given as $A_{\text{total}}(T) = (M_{\text{FeCo}}(0)/\gamma_{\text{FeCo}})(1 - T/T_C)^{\beta_{\text{FeCo}}} - (M_{\text{Gd}}(0)/\gamma_{\text{Gd}})(1 - T/T_C)^{\beta_{\text{Gd}}}$, where $\gamma_{\text{Gd}}$ (or $\gamma_{\text{FeCo}}$) is the gyromagnetic ratio of Gd (or FeCo). The gyromagnetic ratio of Gd (or FeCo) can be defined as $\gamma_{\text{Gd}} = g_{\text{Gd}} \frac{\mu_B}{\hbar}$ (or $\gamma_{\text{FeCo}} = g_{\text{FeCo}} \frac{\mu_B}{\hbar}$), where $g_{\text{Gd}}$ (or $g_{\text{FeCo}}$) is the Landé g factor of Gd (or FeCo), $\mu_B$ is the Bohr magneton and $\hbar$ is the reduced Plank's constant. As $A_{\text{total}} = 0$ at $T = T_A$, the following equation is obtained.

$$T_C - T_A = T_C[(M_{\text{Gd}}(0)g_{\text{FeCo}})/(M_{\text{FeCo}}(0)g_{\text{Gd}})]^{\frac{1}{(\beta_{\text{FeCo}} - \beta_{\text{Gd}})}}. \tag{2}$$

By subtracting Eq. (1) from Eq. (2), we can write the relationship between $T_A$ and $T_M$ as follows.

$$T_A = T_M + T_C\left[1 - (g_{\text{FeCo}}/g_{\text{Gd}})^{\frac{1}{(\beta_{\text{FeCo}} - \beta_{\text{Gd}})}}\right]\left(M_{\text{Gd}}(0)/M_{\text{FeCo}}(0)\right)^{\frac{1}{(\beta_{\text{FeCo}} - \beta_{\text{Gd}})}}. \tag{3}$$

Because of the spin-orbit coupling of FeCo and zero orbital angular momentum of Gd, it is known that $g_{\text{FeCo}}$ ($\sim 2.2$) is slightly greater than $g_{\text{Gd}}$ ($\sim 2$) [30–32]. Consequently, we can expect that $T_A > T_M$ due to the condition of $\beta_{\text{Gd}} > \beta_{\text{FeCo}}$, $M_{\text{Gd}}(0) > M_{\text{FeCo}}(0)$, and $g_{\text{FeCo}} > g_{\text{Gd}}$ [27, 28, 30–32]. From Eq. (3), the linearity of $(T_M, T_A)$ can be easily understood. It is noteworthy that $T_A$ depends on $T_C$ in Eq. (3), and therefore, $T_C$ affects $\Delta T$. This results in a slight deviation from linearity. A standard scaling treatment is employed to examine the universal behaviors. By scaling Eq. (3) and dividing it by $T_C$, we can obtain the relation $T_A/T_C = T_M/T_C + \eta$, where

$$\eta \equiv \left[1 - (g_{\text{FeCo}}/g_{\text{Gd}})^{\frac{1}{(\beta_{\text{FeCo}} - \beta_{\text{Gd}})}}\right]\left(M_{\text{Gd}}(0)/M_{\text{FeCo}}(0)\right)^{\frac{1}{(\beta_{\text{FeCo}} - \beta_{\text{Gd}})}}. \tag{4}$$

As $\eta$ is decided by the given material parameters, Eq. (4) shows that $T_A/T_C$ is directly proportional to $T_M/T_C$.

To confirm the above theoretical prediction, we measured $T_C$ for each sample, as listed in Table II by performing the temperature dependence of the $R_H$ [33]. Figure 4(**a**) shows the variation in $T_A/T_C$ with respect to $T_M/T_C$. This relationship is clearly linear. The slope and y-axis intercept of the best linear fit are 0.99 ± 0.06 and 0.19 ± 0.02, respectively, implying that $\eta$ is approximately constant for the samples with different $T_C$ and $T_A$. Figure 4(**b**) shows the results of $\eta$ for each sample. This result confirms that $\eta$ is invariant (= 0.19) irrespective of the samples. Therefore, the results prove the validity of the general assumption used in the theory.

For a better insight, we study $\eta$ for each parameter: $M_{Gd}(0)$, $M_{FeCo}(0)$, $\beta_{Gd}$, and $\beta_{FeCo}$. Previous studies show that $M_{Gd}(0)/M_{FeCo}(0)$ ranges from 1.1 to 1.2, and $\beta_{Gd}$ and $\beta_{FeCo}$ are 0.45< $\beta_{FeCo}$ <0.5 and 0.65< $\beta_{Gd}$ <0.7, respectively [27, 28]. Using these values, we can numerically calculate $\eta$. The blue dashed lines, shown in Fig. 4(**b**), indicate the calculation results of the upper and lower limits of $\eta$ on the basis of the reported parameters [27, 28]. For the case of typical ranges, the figure shows that the experimental results are in good agreement with the numerical calculation. Therefore, $\eta$ is approximately constant within the experimental accuracy.

From the relation $T_A/T_C = T_M/T_C + \eta$, we can estimate the angular momentum compensation temperature. We denoted the estimated angular momentum compensation temperature as $T_A^*$, which can be calculated using the relation $T_M + \eta T_C$. Here, we used $\eta = 0.19$, from Fig. 4(**b**). To confirm the accordance with the measured angular momentum compensation temperature $T_A$, $T_A^*$ was calculated for all the samples and is plotted with respect to $T_A$, as shown in Fig. 4(**c**). The solid red line ($T_A^* = T_A$) indicates a good conformity of the estimated angular momentum compensation temperature. This observation proves that the experimental inaccuracy in determining $T_A^*$ is within a few Kelvin, which could be acceptable for estimating $T_A$. In Fig. 4(**c**), the inaccuracy remains lower than approximately 5 K.

Finally, it is worthwhile to discuss the general application of the estimation method (of the $T_A^*$) to other ferrimagnetic materials, e.g., TbFeCo. Because the Landé g factors of RE and TM elements, $g_{RE}$ and $g_{TM}$, are different for other elements, $\eta$ can be changed depending on the type of ferrimagnetic material. If $\eta$ can be determined for a given RE–TM ferrimagnet, $T_A^*$ can be easily estimated for any ferrimagnetic material by determining $T_c$ and $T_M$.

In conclusion, we investigate the correlation between $T_M$ and $T_A$ in GdFeCo ferrimagnets. The results show a strong correlation between $T_A$ and $T_M$, which can be demonstrated experimentally and theoretically. Moreover, simple yet efficient method was employed for estimating $T_A$ by measuring $T_C$ and $T_M$. Therefore, this observation will help in easily determining the angular momentum compensation temperature. Accordingly, the proposed scheme can be potentially applied to for ferrimagnet-based spintronics devices.

**Figure Captions**

**Figure 1 (a)** Schematic of the GdFeCo microwire device, and **(b)** Anomalous Hall effect resistance $R_H$ as a function of the perpendicular magnetic field $\mu_0 H_z$ at room temperature (300 K). The orange arrow indicates the coercive field $\mu_0 H_c$ and the black up–down arrow indicates the Hall resistance difference $\Delta R_H \equiv R_H(+H_{z,\text{sat}}) - R_H(-H_{z,\text{sat}})$, where $+H_{z,\text{sat}}$ and $-H_{z,\text{sat}}$ are the saturation fields with the condition $H_c < H_{z,\text{sat}}$.

**Figure 2 (a)** $\Delta R_H$ as a function of the temperature $T$ for Sample II. The blue dot indicates the magnetization compensation temperature $T_M$, and **(b)** DW speed $v$ as a function of $T$ for Sample II at $\mu_0 H_z = 80$ mT. The blue dot indicates the magnetization compensation temperature $T_M$, and the purple arrow indicates the angular momentum compensation temperature $T_A$.

**Figure 3** $T_A$ with respect to $T_M$. The red line is the best linear fit.

**Figure 4 (a)** $T_A/T_c$ as a function of $T_M/T_c$ (the red line is the best linear fit), **(b)** $\gamma$ with respect to the sample number (the red line indicates $\gamma = 0.19$, and the blue dashed lines indicate the calculation of the upper and lower limits of $\gamma$ based on the reported parameters), and **(c)** $T_A^*$ with respect to $T_A$ (the red line is the best linear fit).

**Acknowledgements**

This work was supported by JSPS KAKENHI (Grant Numbers 15H05702, 26870300, 26870304, 26103002, 26103004, 25220604, and 2604316). Collaborative Research Program of the Institute for Chemical Research, Kyoto University, and R & D project for ICT Key Technology of MEXT from the Japan Society for the Promotion of Science (JSPS). This work was partly supported by The Cooperative Research Project Program of the Research Institute of Electrical Communication, Tohoku University. D.H.K. was supported as an Overseas Researcher under Postdoctoral Fellowship of JSPS (Grant Number P16314). D.Y.K. and S.B.C. were supported by a National Research Foundations of Korea (NRF) grant funded by the Ministry of Science, ICT and Future Planning of Korea (MSIP) (2015R1A2A1A05001698 and 2015M3D1A1070465). K.J.K. was supported by the National Research Foundation of Korea (NRF) grant funded by the Korea Government (MSIP) (No. 2017R1C1B2009686, NRF-2016R1A5A1008184) and by the DGIST R&D Program of the Ministry of Science, ICT and Future Planning (17-BT-02).

**Contributions of Authors**

D.H.K. conceptualized the work. D.H.K. and T.O. supervised the study. Y.F., H.Y., and A.T. prepared the films and Y.H. and T.O. made the devices. Y.H., D.H.K., T.O., T.N., and D.Y.K. conducted the experiments. D.H.K. and Y.H. performed the analysis, and D.H.K. modeled the data. D.H.K., T.O., S.B.C., and K.J.K. wrote the manuscript. All authors discussed the results and commented on the manuscript.

Table I. Summary of the sample structures

| | Sample Structures |
|---|---|
| **Sample I** | 5-nm SiN/20-nm $Gd_{23}Fe_{67.4}Co_{9.6}$/100-nm SiN/Si substrate |
| **Sample II** | 5-nm SiN/30-nm $Gd_{23.5}Fe_{66.9}Co_{9.6}$/100-nm SiN/Si substrate |
| **Sample III** | 5-nm SiN/1-nm Gd/ 5-nm $Gd_{23}Fe_{67.4}Co_{9.6}$/1-nm Gd/100-nm SiN/Si substrate |
| **Sample IV** | 5-nm SiN/30-nm $Gd_{23}Fe_{67.4}Co_{9.6}$/5-nm Cu/5-nm SiN/Si substrate |
| **Sample V** | 5-nm SiN/30-nm $Gd_{23.5}Fe_{66.9}Co_{9.6}$/5-nm SiN/Si substrate |
| **Sample VI** | 5-nm SiN/20-nm $Gd_{23}Fe_{67.3}Co_{9.7}$/5-nm Pt/100-nm SiN/Si substrate |

Table II. Curie temperature $T_C$ for each sample

(units: K)

| | Sample I | Sample II | Sample III | Sample IV | Sample V | Sample VI |
|---|---|---|---|---|---|---|
| $T_C$ | $475 \pm 10$ | $450 \pm 10$ | $355 \pm 10$ | $412 \pm 10$ | $412 \pm 10$ | $441 \pm 10$ |

**a**

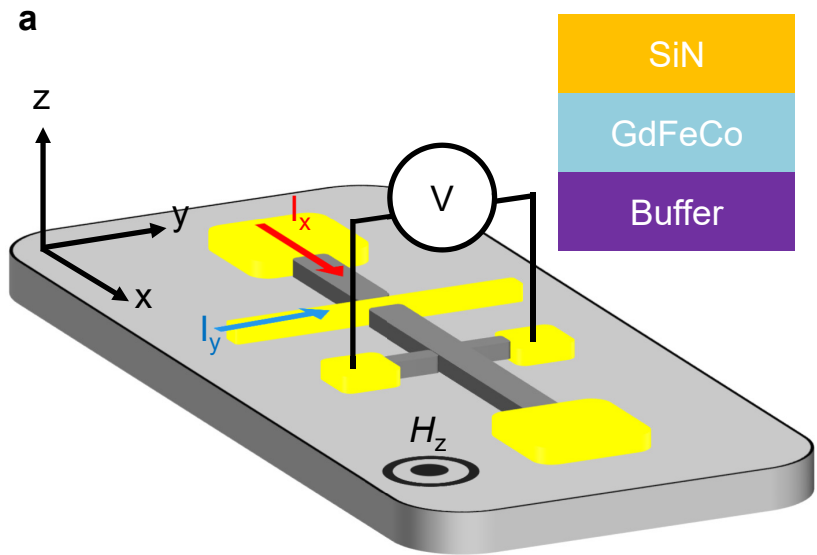

**b**

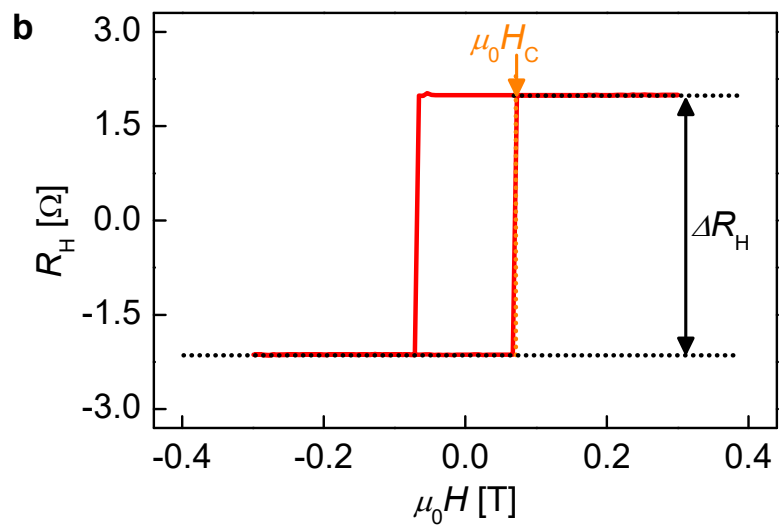

Figure 1

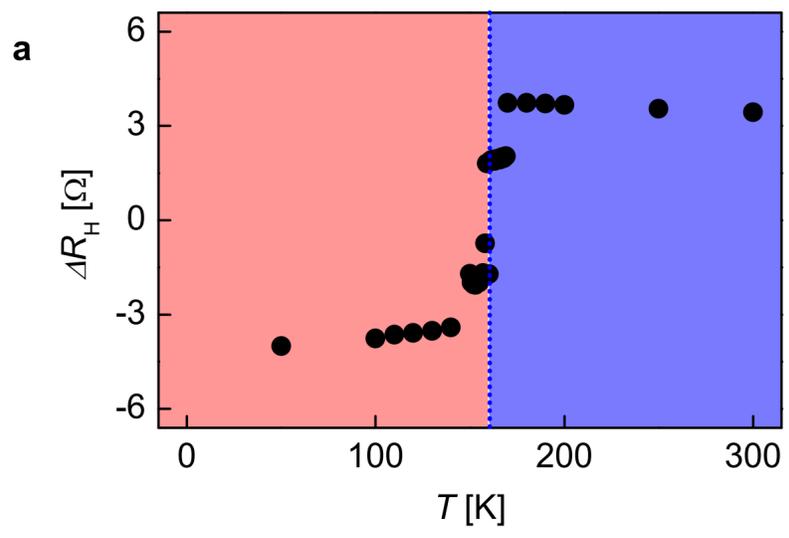

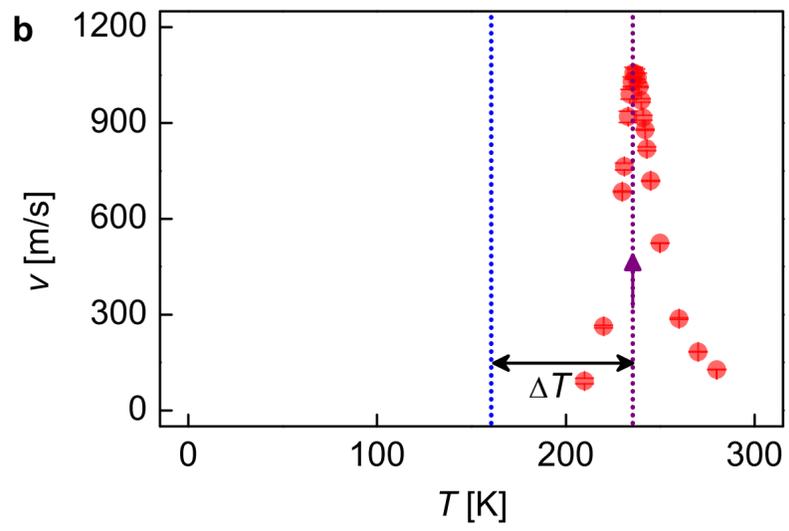

Figure 2

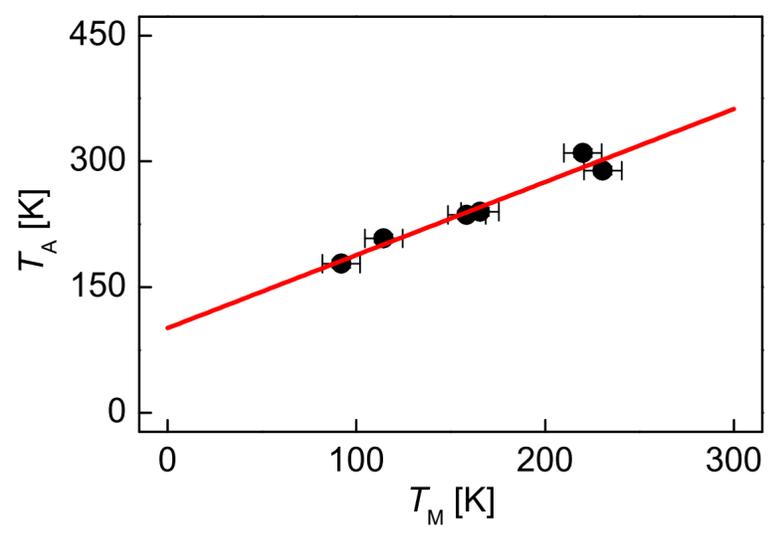

Figure 3

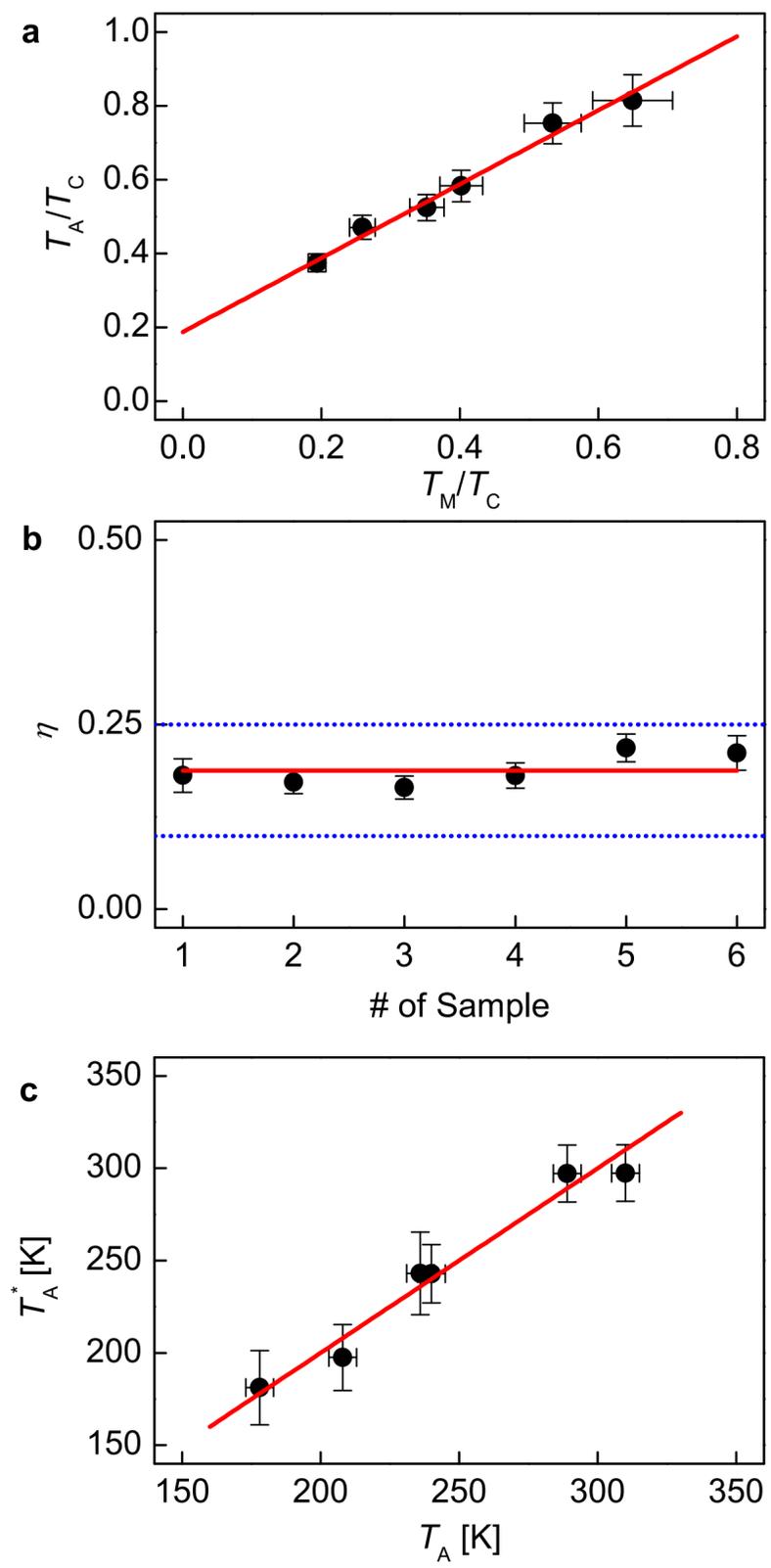

**Figure 4**